\documentclass[journal,a4paper,twocolumn]{IEEEtran}

\usepackage{graphicx}
\usepackage{verbatim}
\usepackage{amssymb,amsmath}
\usepackage{color}
\usepackage{enumerate}

\title{A Physical-layer Rateless Code for Wireless Channels}
\author{Shuang Tian, \textit{Member}, \textit{IEEE}, Yonghui Li, \textit{Senior Member}, \textit{IEEE}, Mahyar Shirvanimoghaddam, \textit{Student Member}, \textit{IEEE}, and Branka Vucetic, \textit{Fellow}, \textit{IEEE}
\thanks{ The authors are with the Telecommunications Laboratory, School of Electrical and Information Engineering,
The University of Sydney, Australia (Email: shuang.tian, yonghui.li, mahyar.shirvanimoghaddam, branka.vucetic@sydney.edu.au).}
\thanks{ This work was supported by the Australian Research Council
(ARC) Grants DP120100190, FT120100487, LP0991663, DP0877090.}}

\begin{document}

\maketitle

\begin{abstract}

In this paper, we propose a physical-layer rateless code for
wireless channels. A novel rateless encoding scheme is developed to
overcome the high error floor problem caused by the low-density
generator matrix (LDGM)-like encoding scheme in conventional
rateless codes. This is achieved by providing each symbol with
approximately equal protection in the encoding process. An extrinsic
information transfer (EXIT) chart based optimization approach is
proposed to obtain a \textit{robust} check node degree distribution,
which can achieve near-capacity performances for a wide range of
signal to noise ratios (SNR). Simulation results show that, under
the same channel conditions and transmission overheads, the
bit-error-rate (BER) performance of the proposed scheme considerably
outperforms the existing rateless codes in additive white Gaussian
noise (AWGN) channels, particularly at low BER regions.

\end{abstract}

\begin{IEEEkeywords}
rateless codes, low-density parity-check (LDPC) codes,
physical-layer channel coding
\end{IEEEkeywords}

\section{Introduction}

\IEEEPARstart{R}{ateless} codes [1] were initially
developed to achieve efficient transmission in erasure channels. The
Luby transform (LT) code [2] is broadly viewed as the
first practical realization of rateless codes. The LT encoded
symbols are generated in accordance with a specific degree
distribution and there is potentially an unlimited number of encoded
symbols. These symbols are generated on the fly and broadcasted to
the receiver until an acknowledgment (ACK) response of successful
decoding is received at the transmitter.

Ideally, a rateless code should allow the receiver to recover the
message symbols with high probability by receiving as few encoded
symbols as possible. Aiming to reduce the necessary number of
encoded symbols, Shokrollahi proposed Raptor codes in [3].
In Raptor codes, a precoding (e.g., based on low-density
parity-check (LDPC) codes) is applied to the message symbols prior
to the LT code. In [3], Raptor codes are constructed in
systematic and non-systematic manners for different erasure channel
scenarios.

The initial work on rateless codes has mainly been limited to
erasure channels with the primary application in multimedia video
streaming. Recently, the design of rateless codes for wireless
channels, such as binary symmetric channels (BSC) [4],
additive white Gaussian noise (AWGN)
channels [5], [6], as well as fading
channels [7], has attracted significant attention.
Rateless codes have a wide spectrum of applications in various
modern wireless communication networks, such as to enhance the
transmission efficiency in IEEE \textit{ad-hoc} 802.11b wireless
networks [8] and cooperative relay
networks [9], [10], and to control the peak-to-average
power ratio in OFDM systems [11]. Rateless codes are
particularly beneficial to wireless transmissions because, in
contrast to traditional fixed-rate coding schemes, the transmitter
potentially does not need to know the channel state information
before sending its encoded symbols, and the receiver can retain a
resilient decoding performance. This property is of particular
importance in the design of codes for time varying wireless
channels.

In general, there are two categories of physical-layer rateless
codes for wireless channels:
systematic [8], [12] and non-systematic [6], [13], [14], [15]. For
systematic rateless codes, the systematic message symbols are first
transmitted to the receiver, followed by a number of encoded
symbols. For non-systematic rateless codes, only encoded symbols are
broadcasted. The receiver uses the classic belief propagation (BP)
algorithm for decoding [16]. At the transmitter, the
encoder of the existing rateless codes uses coding schemes that are
akin to the low-density generator matrix (LDGM) codes [17].
The LDGM-like rateless codes are of linear complexity for encoding
and their encoded symbols can be simply calculated on the fly.

A major drawback of the existing physical-layer rateless codes is
the notably high error floor. This error floor is present mainly
because, from the BP decoding perspective, the encoder creates an
encoded symbol by selecting systematic message symbols only. As a
result, the encoded symbol is connected to only one check node and
cannot receive updated information as the decoding progresses.
Therefore, an erroneous encoded symbol keeps propagating the error
message through its corresponding check node to the systematic
message symbols, which are connected to that check node. To improve
the error floor performance, a considerable amount of research has
been conducted in the past several years, which can be broadly
divided into two directions. One direction applies serial
concatenated coding schemes [13], [14], [18], where the
LDGM-like rateless code is used as a component code in the
concatenation structure. The other direction modifies the LDGM-like
rateless code as a stand-alone code without
concatenation [6], [15]. As reported
in [6], the error performance of concatenation codes is
largely determined by the stand-alone rateless code itself, and the
concatenation will inevitably increase the system complexity.
Therefore, in this paper, we will focus on the design of the
stand-alone rateless code. In [15] and [19], it
shows that the error floor in LDGM-like codes is closely related to
the systematic message symbols connected to a relatively small
number of check nodes. The error floor can be reduced to a certain
level by reducing the number of message symbols of a lower-degree.
For this purpose, [15] proposed to connect all message
symbols to an approximately equal number of check nodes. However,
the method in [15] suffers from a high error floor problem
especially at low signal to noise ratio (SNR) regime. This is
because all encoded symbols still have degree-one, and thus cannot
receive updated information from other check nodes in the BP
decoder.

Another major drawback of the existing physical-layer rateless codes
is that different wireless channel conditions need different check
node degree distributions to achieve the near-capacity
performance [6]. This is primarily because, for the
LDGM-like rateless codes, the degree of systematic message symbols
continues to increase as the transmission goes on to generate and
transmit more encoded symbols. From the extrinsic information
transfer (EXIT) chart perspective [20], increasing the
degree of systematic message symbols will change the EXIT curve of
variable nodes. To achieve a low bit-error-rate (BER) for a near-capacity
transmission overhead, an open but minimum spacing EXIT
chart tunnel is required. As a result, the inverted EXIT curve of
check nodes needs to follow the change, which in turn causes the
check node degree distribution to change. Obviously, to utilize
multiple distributions and adjust them to fit different channel
conditions would complicate the transmission mechanism. A robust
check node degree distribution that can perform well over a wide
range of SNRs would dramatically simplify the system. Some initial
studies [14], [21] have attempted to design a
robust degree distribution, but the progress has been limited so
far.

In this paper, we first develop a rateless encoding approach to
resolve the high error floor problem caused by the conventional
LDGM-like rateless encoding schemes in wireless channels. For the
proposed approach, the encoder creates an encoded symbol by
selecting not only the systematic message symbols, but also the
encoded symbols created prior to the present encoded symbol. By
doing so, each encoded symbol will be connected to multiple check
nodes in the Tanner graph [22]. Moreover, to provide all
the symbols with approximately equal protection in the BP decoder,
we require each variable node to connect to an approximately equal
number of check nodes. To achieve this, we let the variable nodes
with a lower degree be selected with a higher priority when the
encoder selects variable nodes for each check equation to form an
encoded symbol. By using the proposed encoding approach, the encoded
symbol can exchange information with other symbols through multiple
check nodes as the decoding progresses. As a result, the error floor
is lowered.

To provide satisfactory performance over a wide range of SNRs, we
develop an EXIT-chart-based optimization approach and achieve a
near optimal \textit{robust} check node degree distribution for the
proposed rateless code. This robust distribution is able to create a
constant EXIT chart tunnel for different SNRs. In general, for the
conventional rateless codes under a particular channel condition,
the shape of the EXIT curve of variable nodes is determined by the
variable node degree distribution. This distribution depends on two
factors: one is the check node degree distribution, and the other is
the number of symbols transmitted [6]. To achieve a
satisfactory decoding performance, the lower the SNR is, the more
symbols need to be transmitted. Therefore, when SNR changes, the
EXIT curve inevitably changes. In contrast, for the proposed
rateless code, we can keep the EXIT curve unchanged by dividing the
variable nodes involved in the BP decoder into two sets and
manipulating the sizes of the two sets. One set corresponds to the
received symbol sequence from the wireless channel. This received
sequence consists of all the encoded symbols and the message symbols
that are actually transmitted. The other set of variable nodes
corresponds to the remaining message symbols that are yet to be
transmitted. We adjust the number of message symbols in the two sets
in order to achieve a stable EXIT curve of variable nodes for a
desired range of SNRs. Consequently, we can achieve a well-behaved
and constant EXIT chart tunnel leading to superior performances.
Simulation results show that, under the same channel condition and
transmission overhead, the BER performance of the proposed code
considerably outperforms the existing rateless codes in additive
white Gaussian noise (AWGN) channels, particularly at low BER
regions.

The rest of the paper is organized as follows. In Section II, the
system model is introduced. In Section III, the conventional
rateless codes are discussed. In Section IV, the construction of the
proposed rateless code is described first, followed by a discussion
of optimizing the check node degree distribution based on the EXIT
chart analysis. Simulation results are presented in Section V. Some
concluding remarks are made in Section VI.

\section{System Model}

\begin{figure}
\begin{center}
\scalebox{.25}[.25]{\includegraphics{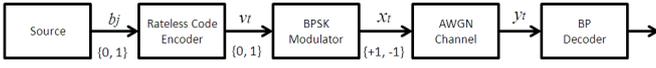}}
\end{center}
\caption{The considered digital communication system model}
\label{Model}
\end{figure}

In this paper, we consider a point-to-point physical-layer
transmission in AWGN channels. The system model is shown in
Fig.~\ref{Model}. Let $ \textbf{b} = \lbrace b_1,...,b_j,...,b_K
\rbrace $ denote the binary message symbols, generated by the
source, where $b_j$ is the $j$-th symbol and $K$ is the length of
message. The message symbols are first encoded by a rateless channel
encoder. Let $v_t$ represent the $t$-th binary rateless coded
symbol, which is then modulated and transmitted. We consider the
binary phase shift keying (BPSK) modulation. Let $x_t \in \{ +1, -1
\}$ be the modulated signal of $v_t$. The received signal, $y_t$, at
the receiver is given by
\begin{equation}
y_t = \sqrt{P_{T}} x_t + n_t \label{y_t}
\end{equation}
\nolinebreak where $P_{T}$ is the transmission power, which is
assumed to be of unity power, i.e. $P_{T} = 1$, and $n_t$ is the
real-valued additive noise in AWGN channels with zero-mean and
variance of $\sigma_n^2$. After receiving the rateless coded
symbols, the receiver applies the BP decoding
algorithm [16] to recover the source message. For the
rateless code, the encoder generates an unlimited number of symbols
and keeps transmitting these symbols until it receives an ACK
response from the receiver.

\section{Conventional Physical-layer Rateless Codes}

\begin{figure}
\begin{center}
\scalebox{.25}[.25]{\includegraphics{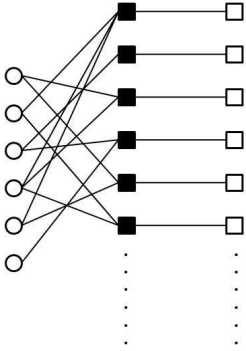}}
\end{center}
\caption{The Tanner graph representing an instance of a systematic
LDGM code [17]. `$\bigcirc$' represents the systematic
message symbols, `$\square$' the encoded symbols and
`$\blacksquare$' the check nodes.} \label{LDGM}
\end{figure}

In the encoding phase, the encoder first draws a value of $d$ from a
distribution called the check node degree distribution. The encoder
then randomly selects $d$ systematic message symbols, and performs
binary summation (i.e. XOR) of these selected symbols to generate an
encoded symbol. The relationship of systematic message symbols and
encoded symbols can be represented by a bipartite Tanner
graph [22]. Fig.~\ref{LDGM} shows the Tanner graph of the
conventional LDGM-like rateless codes. In the Tanner graph, the
systematic message symbols are denoted by `$\bigcirc$' and the
encoded symbols by `$\square$'. The vertices `$\blacksquare$' are
check nodes representing the parity-check equation constraint.

At the receiver, to evaluate the BER performance of the rateless
codes in AWGN channels, Etesami and Shokrollahi introduced a formal
definition of the reception overhead $\delta$ of a decoder
in [6]. Let $M$ denote the number of binary symbols
collected at the receiver to perform satisfactory decoding in the BP
decoder. We express $M$ as
\begin{equation}
M = \frac{K}{C_{\sigma_n}} ( 1 + \delta ) \label{M_reception}
\end{equation}
\nolinebreak where $C_{\sigma_n}$ is the achievable code rate for
binary-input continuous-output AWGN channels with the noise level
parameter $\sigma_n$. This noise level $\sigma_n$ is often called
the \textit{Shannon limit}. From~\eqref{M_reception}, we can see
that the overhead $\delta$ represents the number of collected binary
symbols away from the optimal number defined by $K/C_{\sigma_n}$.

\begin{figure}
\begin{center}
\scalebox{.25}[.25]{\includegraphics{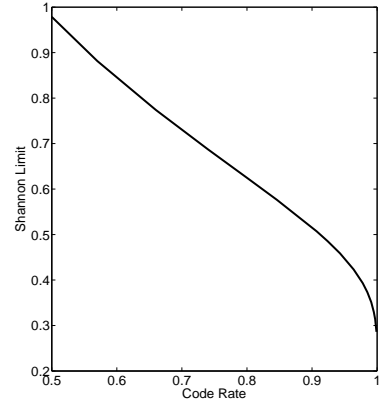}}
\end{center}
\caption{Shannon Limit $\sigma_n$ vs. Code Rate $R$} \label{Shannon}
\end{figure}

For equiprobable $x_t \in \{ +1, -1 \}$, $C_{\sigma_n}$ is given
by [23]
\begin{equation}
C_{\sigma_n} = - \int_{-\infty}^{\infty} \phi_{\sigma_n}(y) \log_2
\phi_{\sigma_n}(y) dy - \frac{1}{2} \log_2 2 \pi e \sigma_n^2
\label{C_Shannon}
\end{equation}
\nolinebreak where
\begin{equation}
\begin{split}
&\phi_{\sigma_n}(y) = \\
&\frac{1}{\sqrt{8 \pi \sigma_n^2 }} \left( \exp \left(
-\frac{(y+1)^2}{2 \sigma_n^2} \right) + \exp \left(
-\frac{(y-1)^2}{2 \sigma_n^2} \right) \right) \, .
\end{split}
\label{Phi_Shannon}
\end{equation}
Since there is no closed form expression, the integral
in~\eqref{C_Shannon} is evaluated by numerical integration
techniques [24]. Table~\ref{Table_Shannon} and
Fig.~\ref{Shannon} show the Shannon limits as a function of some
specific code rates $R$. These values will be used later for the
design of the proposed rateless code. The last column in
Table~\ref{Table_Shannon} is the Shannon limit expressed as the
ratio (dB) of signal energy per symbol to noise power spectral
density, and is defined as
\begin{equation}
\frac{E_b}{N_0} = 10 \log_{10} \left( \frac{1}{2 R \sigma_n^2}
\right) \, . \label{C_Shannon_dB}
\end{equation}

\begin{table*}
\begin{center}
\begin{tabular}{  l | l | l  }
  \hline
  $R$ & $\sigma_n$ & $\frac{E_b}{N_0} \, \textrm{(dB)}$ \\
  \hline
  0.501 & 0.977 & 0.1934 \\
  0.912 & 0.5   & 3.4104 \\
  0.999 & 0.2859 & 7.864
\end{tabular}
\end{center}
\caption{Shannon limits for various code rates $R$.}
\label{Table_Shannon}
\end{table*}

\section{Proposed Physical-layer Rateless Codes Scheme}

In this section, the construction of the proposed rateless code is
described first. A {\it ball-into-bin} method is developed to have
the encoded symbol connected to multiple check nodes, and a full
rank parity-check matrix tailored to the transmission
characteristics of wireless channels is constructed. The
EXIT-chart-based optimization program is then employed to optimize
the robust check node degree distribution.

\subsection{Construction of the Proposed Coding Scheme}

To overcome the high error floor problem in the LDGM-like encoding
scheme, we propose a rateless coding scheme where the encoder
generates an encoded symbol by using not only the systematic message
symbols, but also the encoded symbols created before the present
one. Each encoded symbol will be connected to multiple check nodes
(check equations). Through these check nodes, each variable node and
check node can exchange information with other nodes as the BP
decoding progresses. In this way, the error floor can be lowered
considerably.

\begin{figure}
\begin{center}
\scalebox{.24}[.24]{\includegraphics{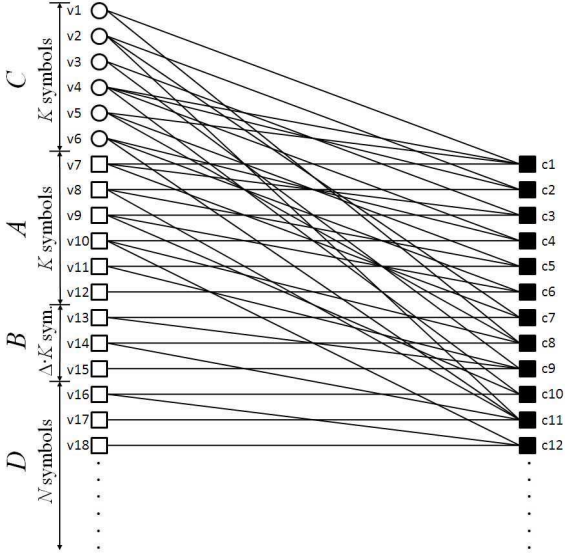}}
\end{center}
\caption{The Tanner graph representing an instance of the proposed
rateless code. `$\bigcirc$' represents the systematic message
symbols, `$\square$' the encoded symbols and `$\blacksquare$' the
check nodes.} \label{Tim_Tanner}
\end{figure}

Let us look at a simple example in Fig.~\ref{Tim_Tanner}, which
illustrates the proposed encoding representation by using a Tanner
graph. In this example, the number of systematic message symbols $K
= 6$. The set of vertices on the left is called variable nodes,
representing the systematic message symbols `$\bigcirc$' and the
encoded symbols `$\square$'. The vertices `$\blacksquare$' on the
right are check nodes. Since each encoded symbol is generated by a
check equation, each variable node corresponding to an encoded
symbol is connected to its corresponding check node (check equation)
in the graph.

\begin{figure}
\begin{center}
\scalebox{.36}[.36]{\includegraphics{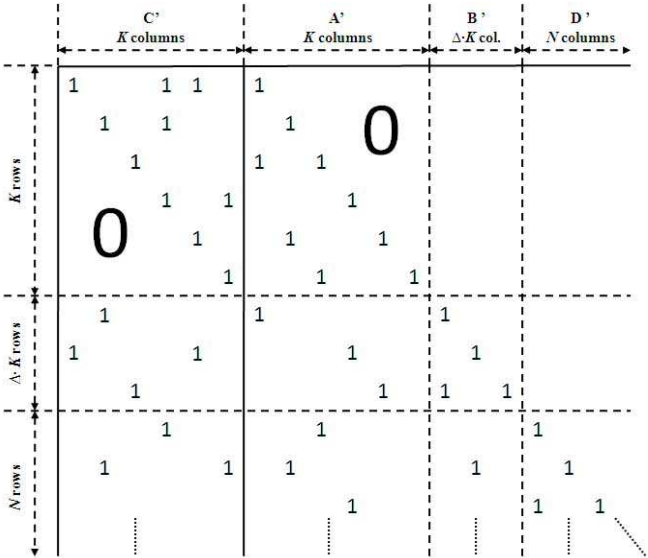}}
\end{center}
\caption{The parity-check matrix representing the instance of the
proposed rateless code in Fig.~\ref{Tim_Tanner}} \label{Tim_matrix}
\end{figure}

An ideal rateless code commonly requires a check node to be randomly
connected to variable nodes. For a check node of degree $i$ in the
LT-based encoding schemes, the check node selects $i-1$ systematic
message symbols in a uniformly random manner to produce an encoded
symbol as shown in Fig.~\ref{LDGM}. This implies that the degree of
the message symbols, or in other words, the number of check nodes a
message symbol connects to, can be very broadly distributed. Some of
the message symbols may not even have a connection to any check node
unless a sufficiently large number of check nodes are provided.
Consequently, even for very high SNRs, there is still a great chance
that the decoding error would occur due to the unconnected message
symbols. Thus, from the transmission efficiency point of view, a
completely random selection would not be suitable for high SNR
regions in AWGN channels. To boost the transmission efficiency at
high SNRs, we apply a full rank constraint to the parity-check
matrix \textbf{H} of the proposed rateless code. The structure of
\textbf{H} of the code in Fig.~\ref{Tim_Tanner} is shown in
Fig.~\ref{Tim_matrix}. Each row of \textbf{H} corresponds to a
parity-check equation and each column of \textbf{H} corresponds to a
symbol in the rateless code. We decompose \textbf{H} into two parts.
The first part of \textbf{H} corresponds to the encoded symbols and
consists of the ($K+1$)-th to the rightmost columns of \textbf{H}.
This part is a sparse lower triangular matrix. The second part of
\textbf{H} corresponds to the $K$ message symbols and consists of
the first $K$ columns of \textbf{H}. In the second part, a sparse
upper triangular matrix is on top of a regular sparse matrix. The
purpose of producing the upper triangular matrix is to guarantee a
full rank matrix for the first $K$ rows of \textbf{H}. This full
rank design ensures that when SNR is sufficiently high, even though
none of the message symbols are transmitted, all message symbols can
be successfully decoded as soon as the decoder receives the first
$K$ encoded symbols. When SNR is lowered, more symbols have to be
transmitted to achieve satisfactory decoding performance. In this
case, we require all the variable nodes to connect to an
approximately equal number of check nodes, which ensures that all
transmitted symbols receive almost the same protection from the BP
decoder. In the following, a \textit{ball-into-bin} encoding scheme
is developed to construct a rateless code as discussed above.

\begin{figure}
\begin{center}
\scalebox{.2}[.2]{\includegraphics{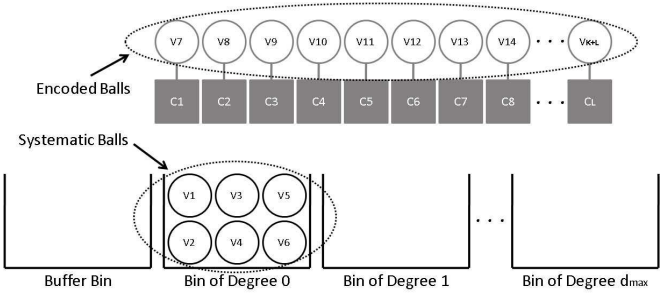}}
\end{center}
\caption{An instance of the \textit{ball-into-bin} encoding process
for $l=0$. `$\bigcirc$' represents the variable balls and
`$\blacksquare$' the check nodes. ($K=6$)} \label{Time_zero}
\end{figure}

\textbf{$\blacklozenge$ \textit{Ball-into-bin} encoding procedure}:
to better describe the encoding process, let us first look at an
example with $K=6$. Suppose there are $K+L$ `variable balls', $v_1,
\dots, v_{K+L}$, corresponding to $K+L$ variable nodes. As shown in
Fig.~\ref{Time_zero}, these variable balls are classified into two
groups. The first group, which we call the \textbf{systematic ball
group}, includes balls, $v_1, \dots, v_K$, representing the $K$
systematic binary message symbols. The second group, which we call
the \textbf{encoded ball group}, contains balls, $v_{K+1}, \dots,
v_{K+L}$, representing the $L$ encoded symbols, and the values for
these encoded symbols are to be determined in the encoding process.
Let $\Omega_i$ denote the check node degree distribution, where $i =
1,2, \dots, i_{max}$. $\Omega_i$ represents the probability that a
check node has degree $i$, and $i_{max}$ is the maximum check node
degree. Similarly, we denote by $d_{max}$ the maximum variable node
degree. There are $d_{max} + 1$ degree bins labeled as `Bin of
Degree $0$', `Bin of Degree $1$', `Bin of Degree $2$', \dots, `Bin
of Degree $d_{max}$'. A ball $v_j$ sitting in `Bin of Degree $d$'
means that $v_j$ is of degree $d$ and has connections to $d$ check
nodes. We also introduce a bin labeled as `Buffer Bin' to produce
the upper triangular matrix in \textbf{H}. Only systematic balls can
be put into `Buffer Bin' by the encoder. If a ball is put into
`Buffer Bin', it will not be used to generate encoded symbols unless
drawn out of `Buffer Bin' and put into the degree bins. The detailed
application of the `Buffer Bin' will be described later in this
section. In addition, since there are $L$ encoded symbols, we have
$L$ check nodes, $c_1, \dots, c_L$ in Fig.~\ref{Time_zero},
correspondingly.

Denoting by $l$ the index of the check nodes, the encoding procedure
progresses as $l$ increases from one to $L$, and is divided into two
phases. \textbf{Phase I}, $1 \le l \le K$, will generate the first
$K$ encoded symbols, which will form the upper triangular matrix of
\textbf{H}; \textbf{Phase II}, $K+1 \le l \le L$, will generate the
rest $L - K$ encoded symbols. Before the encoding process begins
($l=0$, Fig.~\ref{Time_zero}), all systematic balls sit in `Bin of
Degree $0$'.

\begin{figure}
\begin{center}
\scalebox{.2}[.2]{\includegraphics{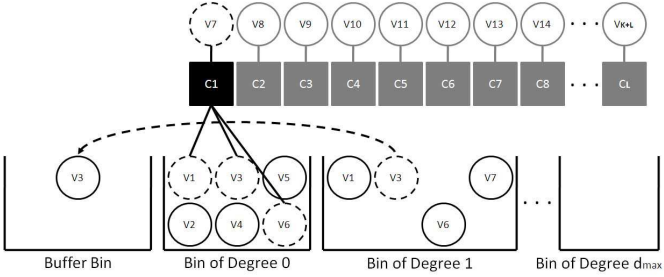}}
\end{center}
\caption{An instance of the \textit{ball-into-bin} encoding process
for $l=1$. `$\bigcirc$' represents the variable balls and
`$\blacksquare$' the check nodes. ($K=6$)} \label{Time_one}
\end{figure}

\textbf{The $1$-st encoding step ($l=1$)}: as shown in
Fig.~\ref{Time_one}, \textit{firstly}, the encoder randomly produces
a degree number $i$ according to the check node degree distribution
$\Omega_i$ and assigns it to the check node $c_1$. In this example,
$i=4$. Since $c_1$ is already connected to $v_7$, the encoder then
picks $i-1$ more variable balls out of `Bin of Degree $0$' in a
uniformly random manner (e.g. $v_1$, $v_3$ and $v_6$) and calculates
the value of $v_7$ by combining the values of these $i-1$ balls
using modulo-2 addition. From the perspective of \textbf{H}, the
first, third, sixth and seventh columns of the first row have
entries of $1$, and the other columns of the same row have entries
of $0$. \textit{Secondly}, the encoder moves the variable balls
involved in the calculation (i.e. $v_1$, $v_3$, $v_6$ and $v_7$) to
`Bin of Degree $1$', since they have had one connection to the check
nodes. \textit{Thirdly}, the encoder randomly chooses one systematic
ball $v_{j^\prime}$ (e.g. $v_3$) from those systematic ones picked
in the present encoding step, and moves it into `Buffer Bin'.
\textit{At last}, the `Buffer Bin' will record the degree bin from
which the systematic ball comes.

\begin{figure}
\begin{center}
\scalebox{.2}[.2]{\includegraphics{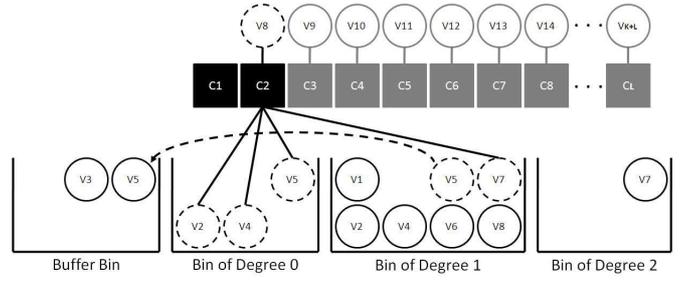}}
\end{center}
\caption{An instance of the \textit{ball-into-bin} encoding process
for $l=2$. `$\bigcirc$' represents the variable balls and
`$\blacksquare$' the check nodes. ($K=6$)} \label{Time_two}
\end{figure}

\textbf{The $2$-nd encoding step ($l=2$)}: as shown in
Fig.~\ref{Time_two}, the encoder randomly produces a degree number
$i$ for $c_2$ (e.g. $i=5$). In order to have each variable node
connected to an approximately equal number of check nodes, we
require the encoder to select as many variable balls as possible in
the bin labeled with the lowest degree. If there are not enough
balls in this bin, the encoder will take all of them and randomly
choose the rest from the bin labeled with the second lowest degree.
It means that balls with a lower degree have a higher priority to be
chosen by the check nodes than those with a greater degree; balls
sitting in the same degree bins, i.e. of the same degree, have the
equal chance to be chosen. In this example, the degree number for
$c_2$ is $i=5$, so the encoder firstly picks all the three variable
balls $v_2$, $v_4$ and $v_5$ in `Bin of Degree $0$' and randomly
chooses a variable ball $v_7$ from `Bin of Degree $1$' to calculate
the value of ball $v_8$. Next, the encoder moves balls $v_2$, $v_4$,
$v_5$ and $v_8$ to `Bin of Degree $1$', and ball $v_7$ to `Bin of
Degree $2$'. Finally, the encoder randomly chooses the systematic
ball $v_5$ and moves it into `Buffer Bin'.

In the remaining of \textbf{Phase I}, a similar encoding procedure
is performed for $3 \le l \le K$. In \textbf{Phase I}, we require
the encoder to pick at least one systematic ball in each encoding
step. Each time $l$ increases by one, the encoder chooses one
systematic ball, which it picks to form the encoded symbol in the
$l$-th encoding step, and moves it into `Buffer Bin'. Systematic
balls sitting in `Buffer Bin' will not be picked again by the
encoder to generate any other encoded symbols until \textbf{Phase I}
ends. It means that, if a systematic ball $v_{j^\prime}$ is put into
`Buffer Bin' in the $l$-th encoding step, the ($l,j^\prime$)-th
entry is `$1$' and there will be no more entries of `$1$' from the
($l+1$)-th to the $K$-th row for the $j^\prime$-th column. By doing
so, as the encoding progresses from the $1$-st to the $K$-th step,
we can obtain a matrix which is equivalent to the upper triangular
sub-matrix of \textbf{H} in Fig.~\ref{Tim_matrix}.

\begin{figure}
\begin{center}
\scalebox{.2}[.2]{\includegraphics{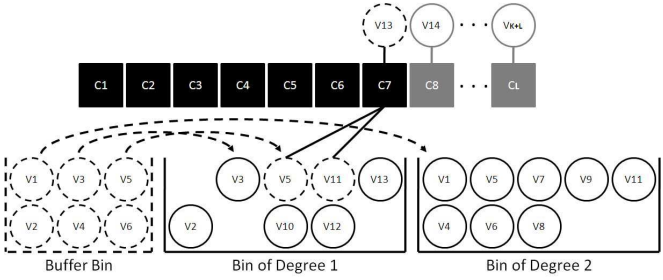}}
\end{center}
\caption{An instance of the \textit{ball-into-bin} encoding process
for $l=7$. `$\bigcirc$' represents the variable balls and
`$\blacksquare$' the check nodes. ($K=6$)} \label{Time_seven}
\end{figure}

\textbf{The ($K+1$)-th encoding step ($l=K+1$)}: as shown in
Fig.~\ref{Time_seven}, \textbf{Phase II} of the encoding process
starts. At this point, the `Buffer Bin' has collected all the $K$
systematic balls. The encoder first puts all these systematic balls
back into those degree bins from which they were moved to the
`Buffer Bin' during \textbf{Phase I}. Next, the encoder performs the
same process as in \textbf{Phase I} except that the `Buffer Bin' is
not used, because the encoder does not move balls to `Buffer Bin' in
\textbf{Phase II}. The encoding process goes on as $l$ increases
until it reaches $L$.

By following the steps above, we obtain a rateless code that can be
represented by a parity-check matrix \textbf{H} as shown in
Fig.~\ref{Tim_matrix}. Next, we discuss the optimization of the
robust check node degree distribution for a desired range of SNRs.

\subsection{Optimization of the Robust Check Node Degree Distribution}

The check node degree distribution is usually expressed in the
polynomial form [6] and is given by
\begin{equation}
\Omega_i(x) = \sum_{i=2}^{i_{max}} \Omega_i x^i \label{Poly_Omega}
\end{equation}
So we have $\Omega_i(1) = 1$. The check node degree distribution can
be defined with respect to the nodes themselves as $\Omega_i(x)$ or
with respect to the edges as $\omega_i(x)$. Let $\omega_i$, $i =
1,2,...,i_{max}$, represent the check node degree distribution with
respect to the edges, i.e. the probability that a given edge is
connected to a check node of degree $i$. The degree polynomial with
respect to the edges is defined by
\begin{equation}
\omega_i(x) = \sum_{i=2}^{i_{max}} \omega_i x^{i-1}
\label{poly_omega}
\end{equation}
The formula for converting between the node-view and edge-view
distributions is given as
\begin{equation}
\omega_i = \frac{\Omega_i \cdot i}{\sum_j \Omega \cdot j} =
\frac{\Omega_i \cdot i }{\beta} \label{omega_convert}
\end{equation}
where $\beta$ is the average degree of check nodes.

Degree distributions are similarly defined for the variable nodes.
We denote $\Lambda_d(x) = \sum_{d} \Lambda_d x^d$ and $\lambda_d(x)
= \sum_{d} \lambda_d x^{d-1}$ the polynomials for the variable nodes
with respect to the nodes and the edges, respectively. Assuming
there are $L$ encoded symbols, and denoting by $\alpha$ the average
degree of variable nodes, we have
\begin{equation}
\alpha (L+K) = \beta L \label{Average_degree_equation}
\end{equation}
Recall that, for the proposed rateless encoding scheme with a
sufficiently large $L$, the majority of variable nodes have a degree
approaching the average degree $\alpha$.
From~\eqref{Average_degree_equation}, the variable node distribution
can be approximately given by a distribution form for variable nodes
in regular LDPC codes as
\begin{equation}
\Lambda_d(x) \approx x^{\alpha} = x^{ \frac{\beta L}{K+L} }, \quad
\lambda_d(x) \approx x^{ \frac{\beta L}{K+L} - 1 }
\label{Poly_Lambda}
\end{equation}
Alternatively, in practice, to achieve a more accurate variable node
degree distribution, we can use computer simulations by first
generating the variable node degree distribution many times using a
fixed check node degree distribution, and then calculating the
empirical mean.

The EXIT chart [20], [25] is a popular curve
fitting tool widely used to optimize degree distributions. This
method tracks the extrinsic mutual information between the
transmitted BPSK symbols and the value of extrinsic log-likelihood
ratios (L-value) in the BP decoder. Let $q$ denote the $q$-th
iteration of the BP decoder. At the $q$-th iteration, an extrinsic
L-value on an edge outgoing from a variable node $v$ to a check node
$c$ is given by
\begin{equation}
LLR_{vc}^{(q)} = LLR_{ch} + \sum_{c'\neq c} LLR_{c'v}^{(q-1)}
\label{Extrinsic_LLR_vTOc}
\end{equation}
where $LLR_{ch}$ is the L-value of the received symbols that are
determined from the channel measurement. $LLR_{cv}^{(q-1)}$ is the
incoming extrinsic L-value, at the $(q-1)$-th iteration, to the
variable node $v$ from a check node $c$, which is in connection to
this variable node, and is given by
\begin{equation}
LLR_{cv}^{(q-1)} = \ln \frac{1 + \prod_{v'\neq v} \tanh(
\frac{LLR_{v'c}^{(q-1)}}{2}) }{1 - \prod_{v'\neq v} \tanh(
\frac{LLR_{v'c}^{(q-1)}}{2}) } \label{Extrinsic_LLR_cTOv}
\end{equation}
For the initial phase where $q = 0$ and from~\eqref{y_t}, we have
\begin{equation}
LLR_{vc}^{(0)} = LLR_{ch} = \ln \frac{p(y|x=+1)}{p(y|x=-1)} =
\frac{2}{\sigma_n^2}y_t , \label{Channel_LLR}
\end{equation}
where $p(y|x)$ is the conditional probability density function (pdf)
of the signal $y$ at the output of the AWGN channel given the input
signal $x \in \{\pm 1 \}$. Thus, we have the variance of $LLR_{ch}$
as
\begin{equation}
\sigma_{ch}^2 = \frac{4}{\sigma_n^2} \label{variance_Channel_LLR}
\end{equation}

The EXIT chart analysis is carried out under two main assumptions.
Firstly, the code Tanner graph can be represented by a tree
graph [6], [26], [27], so that all
involved random variables are independent. The second assumption is
that the extrinsic L-values passed between nodes have a symmetric
Gaussian distribution for mean $\sigma_{LLR}^2/2$ and variance
$\sigma_{LLR}^2$ [20], [26]. Let function
$J(\sigma_{LLR})$ denote the mutual information $I(x_t;LLR)$ between
the transmitted BPSK symbols and the extrinsic L-values with
variance $\sigma_{LLR}^2$, which can be expressed as
\begin{equation}
\begin{split}
&J \left( \sigma_{LLR} \right) = H(x_t) - H(x_t|LLR)\\
&=1 - \int_{-\infty}^\infty \frac{e^{-(\mu -
\sigma_{LLR}^2/2)^2/2\sigma_{LLR}^2}}{\sqrt{2\pi\sigma_{LLR}^2}}
\cdot \log_2(1+e^{-\mu})d\mu \label{J_function}
\end{split}
\end{equation}
where $H(x)$ is the entropy function. Let $I_A$ be the average
mutual information between the BPSK symbols and the L-values
incoming to variable nodes, the average mutual information outgoing
from variable nodes $I_{V-out}$ is then given by
\begin{equation}
I_{V-out} =  \sum_{d=1}^{d_{max}} \lambda_d \cdot J \left( \sqrt{ (d
- 1)[J^{-1}(I_A)]^2 + \sigma_{ch}^2 } \right) \label{I_Vout}
\end{equation}

On the side of check nodes, assume that $I_E$ is the average mutual
information between the BPSK symbols and the outgoing L-values
calculated by check nodes, the average mutual information incoming
to check nodes $I_{C-in}$ is given by
\begin{equation}
I_{C-in} = 1 - \sum_{i=1}^{i_{max}} \omega_i \cdot J \left(
\frac{J^{-1}(1-I_E)}{\sqrt{i-1}} \right) \label{I_Cin}
\end{equation}
The derivation of~\eqref{I_Vout} and~\eqref{I_Cin} is detailed
in [20]. In the EXIT chart analysis, the sets of variable
nodes and check nodes are respectively referred to as the variable
node decoder (VND) and check node decoder (CND) [20].
These two decoders form the BP decoder. The curve $I_{V-out} \, vs.
\, I_A$, $(0 \le I_A \le 1)$, is referred to as EXIT curve of the
VND, and the curve $I_{C-in} \, vs. \, I_E$, $(0 \le I_E \le 1)$, is
referred to as inverted EXIT curve of the CND. Three requirements
need to be satisfied in order to achieve a near optimal degree
design. Firstly, both the VND and inverted CND curves should be
monotonically increasing functions and reach the $(1,1)$ point on
the EXIT chart. Secondly, the VND curve should be always above the
inverted CND curve. And thirdly, the VND curve has to match the
shape of the inverted CND curve as accurately as possible.
Consequently, the VND and inverted CND curves build an EXIT chart
tunnel area. The EXIT chart analysis views iterative decoding as an
evolution of the mutual information in the EXIT tunnel, starting
from point $(0, I_{V-out}(0))$. When the mutual information equals
one, there are no bit errors.

As the transmission progresses, i.e. the code rate changes, for a
fixed degree distribution of check nodes, the inverted CND curve
will stay unchanged. It is desirable if we can obtain a suitable VND
curve that is unchanged as well, so that a stable EXIT chart tunnel
can be obtained. From~\eqref{I_Vout}, it can be observed that the
shape of the VND curve depends on two elements. The first element is
the variable node degree distribution $\lambda_d$, which is in turn
determined by the check node degree distribution and the number of
encoded symbols. The second element for shaping the VND curve is the
channel parameter $\sigma_{ch}^2$. This parameter $\sigma_{ch}^2$
determines the starting point of the VND curves, $( 0, I_{V-out} (0)
)$, which is also the point that the evolution of the mutual
information starts from in the BP decoder.

We start the design of the robust degree distribution by trying to
maintain the quantity of $I_{V-out} (0)$ unchanged as the
transmission goes on. Let us first consider a scenario where the SNR
is sufficiently high, so that the receiver can successfully decode
all the systematic message symbols as soon as it receives $K$
encoded symbols. The reason for sending encoded symbols at first,
rather than sending message symbols themselves, is that the encoded
symbols can provide variable nodes with protection by using parity
check equations in the BP decoder. In this case, the BP decoder has
$2K$ variable nodes. From Fig.~\ref{Tim_Tanner}, we can observe that
these variable nodes can be divided into two sets. One set consists
of the $(K+1)$-th to $2K$-th variable nodes representing $K$ encoded
symbols. For this set, the variable nodes are initiated by L-values
of the received symbols from the AWGN channel. From~\eqref{I_Vout}
and let $I_A = 0$, the quantity of the initial average mutual
information for these variable nodes is given by
\begin{equation}
I_{V-out}^{NZ} =  J \left( \sigma_{ch} \right)
\label{I_Vout_average0_Encode}
\end{equation}
As shown in [20], $J \left( \sigma_{ch} \right)$ is the
capacity of the channel with $\sigma_{ch}$, which in our case is the
achievable code rate for the AWGN channel with BPSK modulation. For
sufficiently high SNRs, $J \left( \sigma_{ch} \right)$ is close to
one. The other variable node set consists of the $K$ message
symbols. Since the message symbols have not been transmitted yet,
the initial mutual information for these variable nodes is given by
$I_{V-out}^{Z} = 0$. Let $K^\prime$ and $\rho_0$ denote the number
and the percentage, respectively, of the variable nodes that have
zero initial mutual information in the BP decoder. We have
\begin{equation}
\rho_0 = \frac{K^\prime}{K+L} \label{rho_define}
\end{equation}
where $L$ is the number of encoded symbols. For the considered
scenario, $K^\prime = L = K$, so $\rho_0 = 0.5$. The quantity of
$I_{V-out} (0)$ is then given by
\begin{equation}
I_{V-out} (0) =  (1 - \rho_0) \cdot I_{V-out}^{NZ} + \rho_0 \cdot
I_{V-out}^{Z} \approx 0.5 \label{I_Vout_average_at0}
\end{equation}

When SNR is lower, more symbols are necessary to be transmitted to
guarantee a reliable decoding. Let us consider a scenario where SNR
is about $0$dB. From Table~\ref{Table_Shannon}, we know that the
achievable code rate $J \left( \sigma_{ch} \right) \approx 0.5$.
From~\eqref{I_Vout_average_at0}, we observe that it requires a
near-to-zero $\rho_0$ to maintain $I_{V-out} (0) \approx 0.5$. A
near-to-zero $\rho_0$ means that, either the BP decoder receives no
message symbols but an infinitely large number of encoded symbols,
i.e. $K^\prime = K$ and $L \to \infty$ in~\eqref{rho_define}, or the
sender starts to supply message symbols at some point of the
transmission, i.e. to decrease $K^\prime$. The latter approach can
effectively reduce $\rho_0$ to zero. In this paper, we refer to it
as the \textbf{\textit{reverse systematic transmission}}.

Let us represent this proposed \textit{reverse} mechanism by
partitioning the variable nodes into four sub-codes, namely Sub-code
$A$, $B$, $C$ and $D$ as shown in the example in
Fig.~\ref{Tim_Tanner}. Sub-code $C$ contains all the systematic
message symbols represented by the first $K$ variable nodes.
Sub-code $A$ is the first $K$ encoded symbols. The $(K+1)$-th to
$\lfloor K \cdot ( 1 + \Delta) \rfloor$-th encoded symbols
constitute Sub-code $B$, where function $\lfloor x \rfloor$
represents the nearest integer to $x$, and $\Delta$ is a positive
value to be determined in the optimization program. Sub-code $D$
consists of encoded symbols from the $(\lfloor K \cdot ( 1 + \Delta)
\rfloor + 1)$-th encoded one to a potentially unlimited number due
to the rateless property. The encoded symbols in Sub-codes $A$ and
$B$ are firstly transmitted in sequence, followed by the message
symbols in Sub-code $C$, and finally more encoded symbols in
Sub-code $D$ are sent. The sender can transmit symbol-by-symbol.
Alternatively, it can transmit block-by-block and each block
consists of a fixed number of symbols. The transmission will be
terminated when the sender receives an ACK response from the
receiver. Next, we address the problem of optimizing the value of
$\Delta$, together with the design of a robust check node degree
distribution.

Similar to~\eqref{I_Vout_average_at0}, the entire curve of
$I_{V-out} \, vs. \, I_A$, $(0 \le I_A \le 1)$, can be given by
\begin{equation}
\begin{split}
& I_{V-out} = \\
& (1 - \rho_0) \cdot \sum_{d=1}^{d_{max}} \lambda_d \cdot J \left(
\sqrt{ (d - 1)[J^{-1}(I_A)]^2 + \sigma_{ch}^2 }
\right) \\
& + \rho_0 \cdot \sum_{d=1}^{d_{max}} \lambda_d \cdot J \left(
\sqrt{ d - 1 } J^{-1}(I_A) \right)
\end{split}
\label{I_Vout_average0}
\end{equation}
It shows that a tunable $\rho_0$ can adjust the shape of the VND
curves. Thus, it is possible for us to keep the VND curve unchanged
by adjusting $\rho_0$ while the transmission goes on. As shown in
Fig.~\ref{Tim_Tanner}, for the reverse systematic transmission, the
message symbols are transmitted after $ \lfloor K \cdot (1 + \Delta)
\rfloor $ encoded symbols. When all the message symbols are received
at the decoder, we have $K^\prime = 0$, so $\rho_0$ would always be
zero and, consequently, the decoder will lose the ability to adjust
the VND curves. In order to have a tunable $\rho_0$, $\Delta$ is
expected to be large enough so that the decoding can be successful
before all the message symbols are sent. In this case, for a single
SNR value of $\xi$dB, the rate of the proposed rateless code can be
calculated as
\begin{equation}
\begin{split}
R_{\xi dB} & = \frac{K}{(K+K+\Delta_{\xi dB} \cdot K)(1 -
\rho_{0,{\xi dB}})} \\
& = \frac{\alpha_{\xi dB}}{(1+\Delta_{\xi dB})(1-\rho_{0,{\xi dB}})}
\sum_{i=2}^{i_{max}} \frac{\omega_i}{i}
\end{split}
\label{Rate_1}
\end{equation}
It can be observed that the actual code rate depends on the degree
distributions, as well as on $\Delta_{\xi dB}$ and $\rho_{0,{\xi
dB}}$. Based on this observation, an optimization program is
developed to obtain a robust check node degree distribution.

Let the SNR range for the robust distribution target on $[SNR_{low},
SNR_{high}]$. We choose $\tau$ SNR points within this range, where $
SNR_{low} \le SNR_1 < SNR_2 < ... < SNR_{\tau - 1} < SNR_{\tau} \le
SNR_{high}$. Let $C_{SNR_\zeta}$ denote the capacity of code rate
corresponding to $SNR_\zeta$, where $1 \le \zeta \le \tau$. Further,
denote $\chi_{SNR_\zeta}$ as
\begin{equation}
\chi_{SNR_\zeta} = \frac{C_{SNR_\zeta}}{R_{SNR_\zeta}}
\label{Chi_dif}
\end{equation}
where $R_{SNR_\zeta}$ is given by~\eqref{Rate_1}. We want
$\chi_{SNR_\zeta}$ to be as close to one as possible, which means
$R_{SNR_\zeta}$ is close to the capacity. For the given
$\Delta_{SNR_\zeta}$, $\rho_{0,SNR_\zeta}$ and $\alpha_{SNR_\zeta}$,
the robust degree distribution of $\omega_i$ is to minimize the
maximum element of
\begin{equation}
\lbrace \chi_{SNR_1}, \, \chi_{SNR_2},  \, ... ,  \,
\chi_{SNR_{\tau-1}}, \, \chi_{SNR_{\tau}} \rbrace
\label{Opt_function}
\end{equation}
subject to four constraints
\begin{enumerate}
\item $\quad \sum_{i=2}^{i_{max}} \omega_i = 1$,
\item $\quad \omega_i \ge 0$,
\item $\quad |I_{V-out, SNR_\zeta} - I_{V-out, SNR_{\zeta '}}| <
\varepsilon, \, \textrm{where} \, \, \, \varepsilon \to 0$,
\item $\quad I_{V-out, SNR_\zeta} > I_{C-in}$ for $I_A = I_E$.
\end{enumerate}
This optimization program can be solved by using a linear program
with computer search. Condition (3) is used to enforce each pair of
the VND curves being as close to each other as possible, so that a
stable EXIT chart tunnel is obtained. Condition (4) is used to
enforce the condition that the EXIT chart tunnel for every
$SNR_\zeta$ point stays open. Therefore, an arbitrarily low BER can
be achieved with the near-capacity transmission overhead. In addition, to
reduce the transmission complexity, it is ideal to have one global
$\Delta$. We propose to select that value as $\Delta = \max \lbrace
\Delta_{SNR_1}, \Delta_{SNR_2}, ... \Delta_{SNR_{\tau -
1}},\Delta_{SNR_\tau} \rbrace$, covering all $\Delta_{SNR_\zeta}$.
Generally, $\Delta$ is determined by the lowest SNR point, because
the lower the SNR is, the more transmitted symbols are required.
Finally, in practice, we found that it is more efficient to perform
the degree optimization program by dealing with the lower SNR points
first.

\begin{figure}
\begin{center}
\scalebox{.3}[.3]{\includegraphics{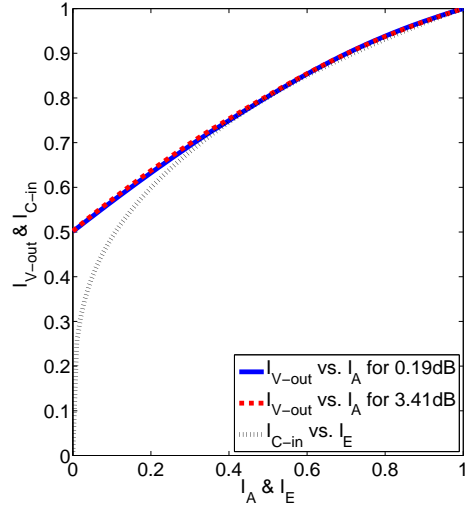}}
\end{center}
\caption{The EXIT chart for $I_{V-out} \, vs. \, I_A$ (curve of VND)
and $I_{C-in} \, vs. \, I_E$ (inverted curve of CND)}
\label{EXIT_chart}
\end{figure}

An example of the EXIT chart resulting from the computer search for
two SNR points $0.19$dB and $3.41$dB is shown in
Fig.~\ref{EXIT_chart}. These two points were chosen to correspond to
a lower code rate $R_{(\sigma_n = 0.977)} = 0.501$ and a higher code
rate $R_{(\sigma_n = 0.5)} = 0.912$, respectively, as shown in
Table~\ref{Table_Shannon}. In this computer search, we found the
values $\Delta = 0.3$, $\rho_{0, \, 0.19dB} = 0$, $\rho_{0, \,
3.41dB} = 0.45$, and the robust check node degree distribution as
\begin{equation}
\Omega(x) = 0.475 x^3 + 0.525 x^6 \label{Opt_Poly_Omega}
\end{equation}
In Fig.~\ref{EXIT_chart}, we can observe that the two respective VND
curves obtained by using the robust check node degree distribution
in~\eqref{Opt_Poly_Omega} coincide with each other closely. It
indicates that the system complexity can be reduced under different
channel conditions.

\section{Simulation Results}

In this section, we will present the BER performance as a function
of the transmission overhead for various rateless coding schemes.
BPSK modulation is used in the simulation and the transmission
signal power is set to unity. We first consider two scenarios in an
AWGN channel, where the noise standard deviations are $\sigma_n =
0.977$ and $0.5$, which correspond to $0.19$ dB and $3.41$ dB,
respectively, as shown in Table~\ref{Table_Shannon}. Given that the
current practical code-length requirement is a few
thousands [28], [29] and with regards to the
necessity of sparseness in LDPC-like codes [30], the number
of message symbols $K$ used in simulations is set to 2000, 3000 and
5000. The comparison is given between the proposed scheme, the
non-systematic rateless code developed in [15], and the
systematic rateless code in [12].

\begin{figure}
\begin{center}
\scalebox{.32}[.32]{\includegraphics{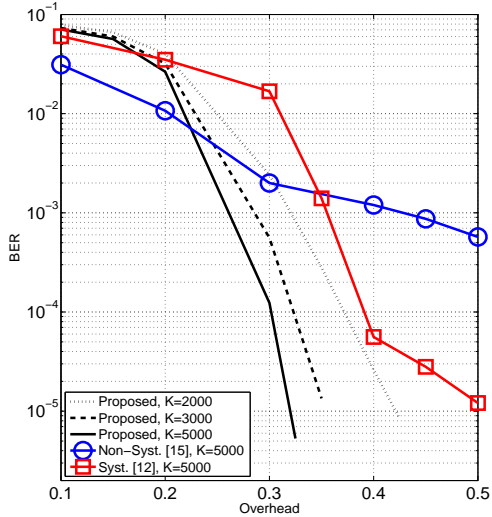}}
\end{center}
\caption{BER vs. Overhead for $\sigma_n = 0.977$} \label{BER_0db}
\end{figure}

Fig.~\ref{BER_0db} shows the BER performance with respect to
transmission overhead $\delta$ at the low SNR region for $\sigma_n =
0.977$. From~\eqref{M_reception} and Table~\ref{Table_Shannon}, the
number of received symbols is given by
\begin{equation}
M = \frac{K}{0.501} (1+\delta) \label{Mbits_0dB}
\end{equation}
We can observe that, for the proposed rateless code at a fixed BER
level, the overhead decreases rapidly as $K$ increases. For
$K=5000$, we can see that the schemes in [15] and [12] have an error floor at BER of $10^{-3}$ and
$10^{-5}$ respectively, while the proposed scheme achieves
significant performance improvement and the error floor has not
appeared at BER of $10^{-5}$. Moreover, for BER of $10^{-5}$, the
proposed scheme saves
\begin{equation}
M' = \frac{5000}{0.501} \times 0.18 \approx 1796
\label{Mbits_0dB_save}
\end{equation}
redundant symbols on average, i.e. almost $36\%$ of $K$, compared to
the existing rateless coding schemes. The improved performance
brought by the proposed rateless coding scheme is the result of the
elimination of the degree-one encoded symbols and the application of
the specially designed full rank parity-check matrix \textbf{H}. We
also compared the performance of these schemes for $K = 2000$ and
$3000$, and obtained similar results. To simplify the illustration
and reduce the number of curves, we only show the comparisons for $K
= 5000$ in the figure.

\begin{figure}
\begin{center}
\scalebox{.32}[.32]{\includegraphics{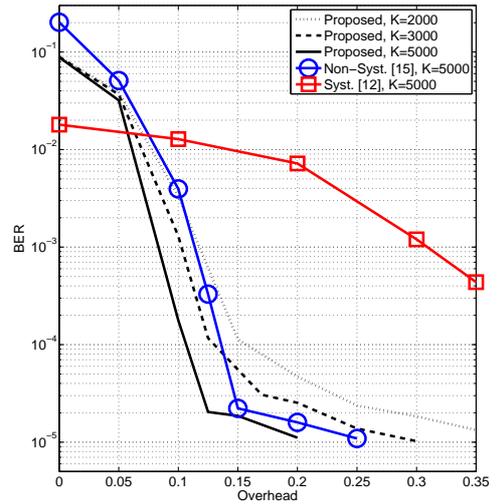}}
\end{center}
\caption{BER vs. Overhead for $\sigma_n = 0.5$} \label{BER_3db}
\end{figure}

Fig.~\ref{BER_3db} compares the BER performance at a higher SNR
region for $\sigma_n = 0.5$. Similarly, for the proposed code at a
fixed BER level, the overhead decreases as $K$ increases. It can
also be observed that the systematic code gives the worst
performance for BER lower than $10^{-2}$. Although both the proposed
and the non-systematic coding schemes have error floors at BER of
about $10^{-5}$, the proposed scheme still outperforms the
non-systematic one and saves about
\begin{equation}
M' = \frac{5000}{0.912} \times 0.05 \approx 274
\label{Mbits_3dB_save}
\end{equation}
redundant symbols on average, i.e. about $5.5\%$ of $K$.

\begin{figure}
\begin{center}
\scalebox{.32}[.32]{\includegraphics{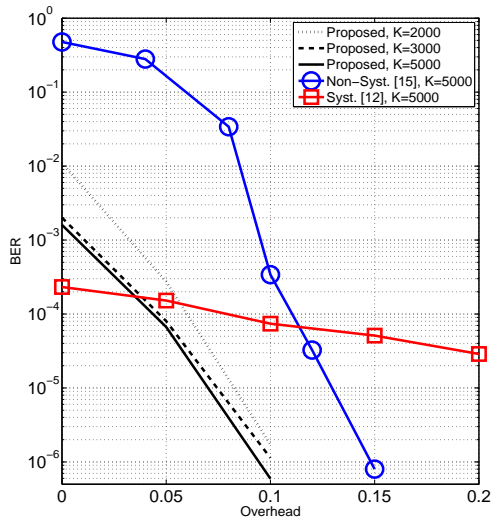}}
\end{center}
\caption{BER vs. Overhead for $\sigma_n = 0.2859$} \label{BER_78db}
\end{figure}

In Fig.~\ref{BER_78db}, the BER performance for $\sigma_n = 0.2859$
is plotted. We can observe that the proposed scheme has the best
performance for BER better than $10^{-4}$. The same robust check
node degree distribution used in Fig.~\ref{BER_0db} and
Fig.~\ref{BER_3db} is also used here. For $K=2000$, BER of $10^{-5}$
is reached with overhead less than $0.1$. The results are fairly
satisfactory, although the case of $\sigma_n = 0.2859$ was not taken
into consideration when we conducted the computer search in the
optimization program in Section IV B.

\begin{table*}
\begin{center}
\begin{tabular}{  l | l | l  }
  \hline
  Index & Degree Distribution $\Omega(x)$ & Threshold $\sigma_{th}$ \\
  \hline
  DD1 & $0.475 x^3 + 0.525 x^6$ & 0.53 \\
  DD2 & $0.1 x^2 + 0.4 x^3 + 0.5 x^6$   & 0.525 \\
  DD3 & $0.6 x^3 + 0.4 x^6$ & 0.44
\end{tabular}
\end{center}
\caption{Three degree distributions and their corresponding
threshold values.} \label{Table_Threshold}
\end{table*}

In addition, we provide the experimental results for different check
node degree distributions to justify the use of EXIT chart analysis
in the proposed encoding method. We first choose three given degree
distributions, and then obtain their channel threshold parameters by
using the EXIT chart approach. Thirdly, we perform simulations for
the decoding performance. Results show that the degree distribution
of a greater threshold does give a better decoding performance. This
is consistent with the EXIT chart analysis.

\begin{figure}
\begin{center}
\scalebox{.32}[.32]{\includegraphics{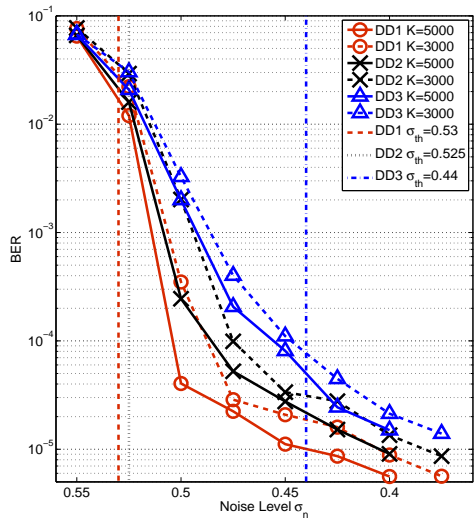}}
\end{center}
\caption{BER vs. Noise Level $\sigma_n$} \label{Threshold_comp}
\end{figure}

We consider a code rate $R=0.8$, and the three degree distributions
and their channel threshold values $\sigma_{th}$ obtained by EXIT
chart analysis are given in Table~\ref{Table_Threshold}. Next, we
perform simulations for the three degree distributions, which are
shown as DD1, DD2 and DD3 in Fig.~\ref{Threshold_comp}, with
code-length of 6250 and 3750 symbols (i.e. the number of message
symbols $K$ is 5000 and 3000 respectively). The threshold value
$\sigma_{th}$ of each degree distribution is also shown in
Fig.~\ref{Threshold_comp}.

From EXIT chart analysis, we know that the threshold values
$\sigma_{th}$ of DD1 and DD2 are 0.53 and 0.525 respectively.
Simulation results show that DD1 and DD2 have waterfall regions that
closely match the threshold prediction by the EXIT chart analysis.
As the code-length goes very long, we can presume an even closer
match in the threshold between the simulations and the EXIT chart
analysis. For DD3, it does not have clear waterfall and error floor
regions, because its threshold is too far away from the Shannon
limit for code rate $R = 0.8$, so DD3 is far from a good
distribution choice for this code rate. Overall, we come to the
conclusion from simulation results that the degree distribution of a
greater threshold has a better decoding performance, so the EXIT
chart analysis is justified in the degree distribution design for
the proposed rateless encoding method.

\section{Conclusions}

In this paper, we propose a physical-layer rateless coding scheme
for wireless channels. A {\it ball-into-bin} method is firstly
introduced to construct a particular parity-check matrix \textbf{H}.
The aim of this method is to let both the systematic message symbols
and the encoded symbols have an approximately equal degree, so that
they can receive equal protection in the BP decoder. Moreover, in order
to obtain a robust degree distribution of check nodes for a range of
SNRs, a degree optimization program based on the EXIT chart analysis
is proposed. The proposed rateless code with the robust degree
distribution achieves near-capacity performances with respect to the
transmission overhead. Simulation results show that the transmission
redundancy is reduced as the message length increases. Moreover,
under the same channel condition and transmission overhead, the BER
performance of the proposed code outperforms the existing rateless
codes in AWGN channels, particularly at low BER regions. For
example, for BER of $10^{-5}$ and $E_b/N_0$ of $0.19$ dB, the
proposed scheme saves more than one-third of the message length
compared to the existing rateless schemes.

\section*{References}
\footnotesize{
\begin{enumerate}
\renewcommand{\labelenumi}{[\theenumi]}
\item J. Byers, M. Luby, M. Mitzenmacher and A. Rege, ``A digital fountain
approach to reliable distribution of bulk data,'' in \textit{Proc.
ACM SIGCOMM}, Vancouver, BC, Canada, Jan. 1998, pp. 56 - 67.
\item M. Luby, ``LT codes,'' in \textit{Proc. The 43rd Annual IEEE Symposium on
Foundations of Computer Science}, Vancouver, BC, Canada, Nov. 2002,
pp. 271 - 280.
\item A. Shokrollahi, ``Raptor codes,'' \textit{IEEE Transactions on Information
Theory}, vol. 52, no. 6, pp. 2551 - 2567, June 2006.
\item R. Palanki and J. Yedidia, ``Rateless codes on noisy channels,'' in
\textit{Proc. IEEE International Symposium on Information Theory
(ISIT)}, Chicago, IL, June 2004, pp. 38.
\item O. Etesami, M. Molkaraie and A. Shokrollahi, ``Raptor Chodes on
Symmetric Channels,'' in \textit{Proc. IEEE International Symposium
on Information Theory (ISIT)}, Chicago, IL, June 2004, pp. 39.
\item O. Etesami and A. Shokrollahi, ``Raptor codes on binary memoryless
symmetric channels,'' \textit{IEEE Transactions on Information
Theory}, vol. 52, no. 5, pp. 2033 - 2051, May 2006.
\item J. Castura and Y. Mao, ``Raptor coding over fading channels,'' \textit{IEEE
Communications Letters}, vol. 10, no. 1, pp. 46 - 48, Jan. 2006.
\item A. Ngo, T. Steven, G. Maunder and L. Hanzo, ``A systematic LT coded
arrangement for transmission over correlated shadow fading channels
in 802.11 ad-hoc wireless networks,'' in \textit{Proc. IEEE
Vehicular Technology Conference (VTC 2010-Spring)}, Taipei, Taiwan,
May 2010, pp. 1 - 5.
\item X. Liu and T. Lim, ``Fountain codes over fading relay channels,'' \textit{IEEE
Transactions on Wireless Communications}, vol. 8, no. 6, pp. 3278 - 3287, June 2009.
\item W. Chen and W. Chen, ``A new rateless coded cooperation scheme for
multiple access channels,'' in \textit{Proc. IEEE International
Conference on Communications (ICC)}, Kyoto, Japan, June 2011, pp. 1 - 5.
\item T. Jiang and X. Li, ``Using fountain codes to control the peak-to-average power ratio of OFDM signals,'' \textit{IEEE Transactions on
Vehicular Technology}, vol. 59, no. 8, pp. 3779 - 3785, Oct. 2010.
\item T. Nguyen, L. Yang and L. Hanzo, ``Systematic Luby transform codes
and their soft decoding,'' in \textit{Proc. IEEE Workshop on Signal
Processing Systems}, Shanghai, China, Oct. 2007, pp. 67 - 72.
\item Z. Cheng, J. Castura and Y. Mao, ``On the design of raptor codes
for binary-input gaussian channels,'' \textit{IEEE Transactions on
Communications}, vol. 57, no. 11, pp. 3269 - 3277, Nov. 2009.
\item R. Barron, C. Lo and J. Shapiro, ``Global design methods for raptor codes
using binary and higher-order modulations,'' in \textit{Proc. IEEE
Military Communications Conference}, Boston, MA, Oct. 2009, pp. 1 - 7.
\item I. Hussain, M. Xiao and L. K. Rasmussen, ``Error floor analysis of LT
codes over the additive white Gaussian noise channel,'' in
\textit{Proc. IEEE Global Telecommunications Conference (GLOBECOM)},
Houston, USA, Dec. 2011, pp. 1 - 5.
\item R. G. Gallager, \textit{Low density parity-check codes}. Cambridge, MA: MIT
Press, 1963.
\item J. Garcia-Frias and W. Zhong, ``Approaching Shannon performance by
iterative decoding of linear codes with low-density generator
matrix,'' \textit{IEEE Communications Letters}, vol. 7, no. 6, pp.
266 - 268, June 2003.
\item N. Bonello, R. Zhang, S. Chen and L. Hanzo, ``Reconfigurable rateless
codes,'' \textit{IEEE Transactions on Wireless Communications}, vol.
8, no. 11, pp. 5592 - 5600, Nov. 2009.
\item K. Liu and J. Garcia-Frias, ``Error floor analysis in LDGM codes,'' in
\textit{Proc. IEEE Int. Symp. Inf. Theory (ISIT)}, Austin, USA, Jun.
2010, pp. 734 - 738.
\item S. ten Brink, G. Kramer and A. Ashikhmin, ``Design of low-density
parity-check codes for modulation and detection,'' \textit{IEEE
Transactions on Communications}, vol. 52, no. 4, pp. 670 - 678,
April 2004.
\item R. Tee, T. Nguyen, L. Yang and L. Hanzo, ``Serially concatenated Luby
Transform coding and bit-interleaved coded modulation using
iterative decoding for the wireless internet,'' in \textit{Proc.
IEEE Vehicular Technology Conference, 2006-Spring}, Melbourne,
Australia, May 2006, pp. 22 - 26.
\item R. Tanner, ``A recursive approach to low complexity codes,'' \textit{IEEE
Transactions on Information Theory}, vol. 27, no. 5, pp. 533 - 547,
Sept. 1981.
\item S. J. Johnson, \textit{Iterative error correction}. Cambridge, United Kingdom:
Cambridge University Press, 2010.
\item S. Lin and D. Costello, \textit{Error control coding: fundamentals and applications}.
Upper Saddle River, NJ, U.S.A.: Pearson Education, Inc., 2nd
Edition, 2004.
\item S. ten Brink, ``Convergence behavior of iteratively decoded parallel
concatenated codes,'' \textit{IEEE Transactions on Communications},
vol. 49, no. 10, pp. 1727 - 1737, Oct. 2001.
\item T. Richardson, A. Shokrollahi and R. Urbanke, ``Design of capacity approaching
irregular low-density parity-check codes,'' \textit{IEEE
Transactions on Information Theory}, vol. 47, no. 2, pp. 619 - 637,
Feb. 2001.
\item M. Luby, M. Mitzenmacher and A. Shokrollahi, ``Analysis of random
processes via and-or tree evaluation,'' in \textit{Proc. 9th Annu.
ACM-SIAM Symp. Discrete Algorithms}, San Francisco, CA, Jan. 1998,
pp. 364 - 373.
\item \textit{IEEE standard for local and metropolitan are networks, Part 16: Air
interface for fixed and mobile broadband wireless access systems}.
IEEE Std. 802.16, 2009.
\item \textit{IEEE 802.11n-2009: Wireless LAN Medium Access Control (MAC)
and Physical Layer (PHY) Specifications: Enhancements for Higher
Throughput}. IEEE Std. 802.11, 2009.
\item D. MacKay, S. Wilson and M. Davey, ``Comparison of construction
of irregular Gallager codes,'' \textit{IEEE Transactions on
Communications}, vol. 47, no. 10, pp. 1449 -
1454, Oct. 1999.

\end{enumerate}
}


\begin{IEEEbiography}
[{\includegraphics[width=1in,height=1.25in,clip,keepaspectratio]{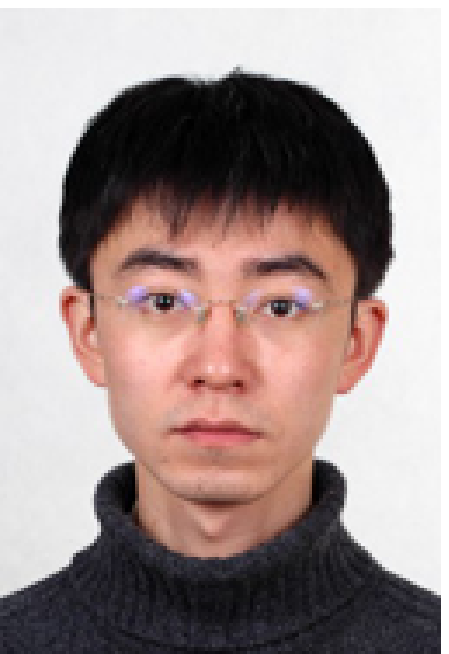}}]{Shuang
Tian} (M'13) received the B. Sc. degree (with highest honors) from
Harbin Institute of Technology, China, in 2005, the M. Eng. Sc. (Research) degree from Monash University, Australia, in 2008, and is
currently working toward the Ph.D. degree at The University of
Sydney, Australia, all in electrical engineering. From 2008 to 2009,
he worked as a system and standard engineer with Huawei Technologies
Co. Ltd., China. His research interests include coding techniques,
cooperative communications, 3GPP/WiMAX network protocols, OFDM
communications, and optimization of resource allocation and interference management
in heterogeneous networks. He is a recipient of University of Sydney Postgraduate Awards.
\end{IEEEbiography}

\begin{IEEEbiography}
[{\includegraphics[width=1in,height=1.5in,clip,keepaspectratio]{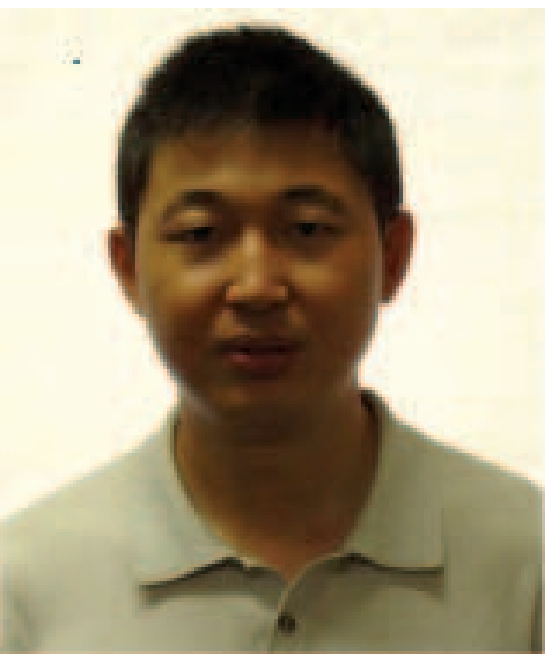}}]{Yonghui
Li} (M'04-SM'09) received his Ph.D. degree in November 2002 from
Beijing University of Aeronautics and Astronautics. From 1999 - 2003, he was affiliated with Linkair Communication Inc, where he
held a position of project manager with responsibility for the
design of physical layer solutions for the LAS-CDMA system. Since
2003, he has been with the Centre of Excellence in
Telecommunications, The University of Sydney, Australia. He is now
an Associate Professor in School of Electrical and Information
Engineering, The University of Sydney. He was the Australian Queen
Elizabeth II Fellow and is currently the Australian Future Fellow.

His current research interests are in the area of wireless
communications, with a particular focus on MIMO, cooperative
communications, coding techniques and wireless sensor networks. He
holds a number of patents granted and pending in these fields. He is
an executive editor for European Transactions on Telecommunications
(ETT). He has also been involved in the technical committee of
several international conferences, such as ICC, Globecom, etc.
\end{IEEEbiography}

\begin{IEEEbiography}
[{\includegraphics[width=1in,height=1.25in,clip,keepaspectratio]{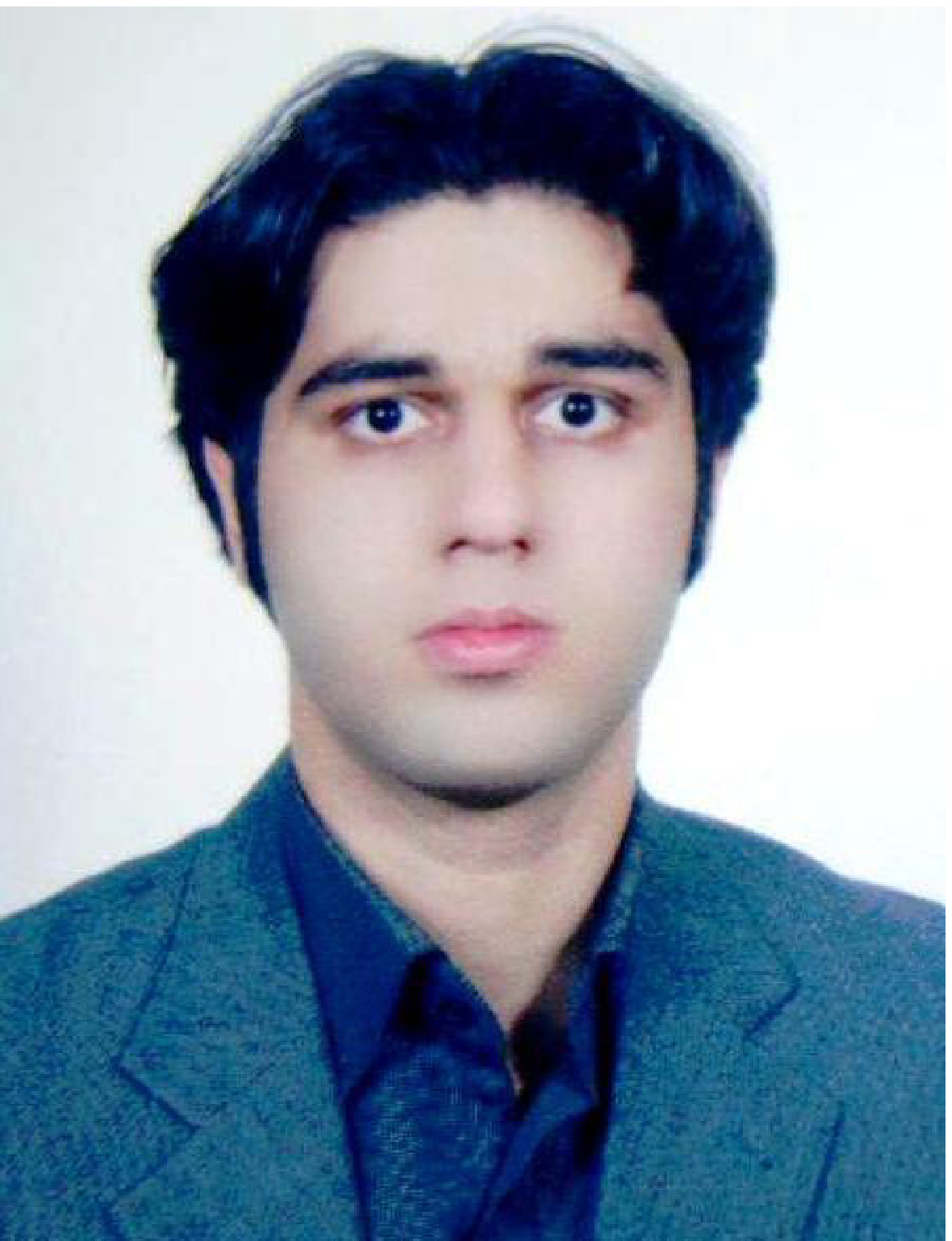}}]{Mahyar Shirvanimoghaddam}
received the B. Sc. degree with the 1'st Class Honour from Tehran University, Iran, in 2008, and the M. Sc. Degree with the 1'st Class Honour from Sharif University of Technology, Iran, in 2010, and is currently working toward the Ph.D. degree at The University of Sydney, Australia, all in Electrical Engineering. His research interests include coding techniques, cooperative communications, compressive sensing, and wireless sensor networks. He is a recipient of University of Sydney International Scholarship (USydIS) and University of Sydney Postgraduate Awards.
\end{IEEEbiography}

\begin{IEEEbiography}
[{\includegraphics[width=1in,height=1.25in,clip,keepaspectratio]{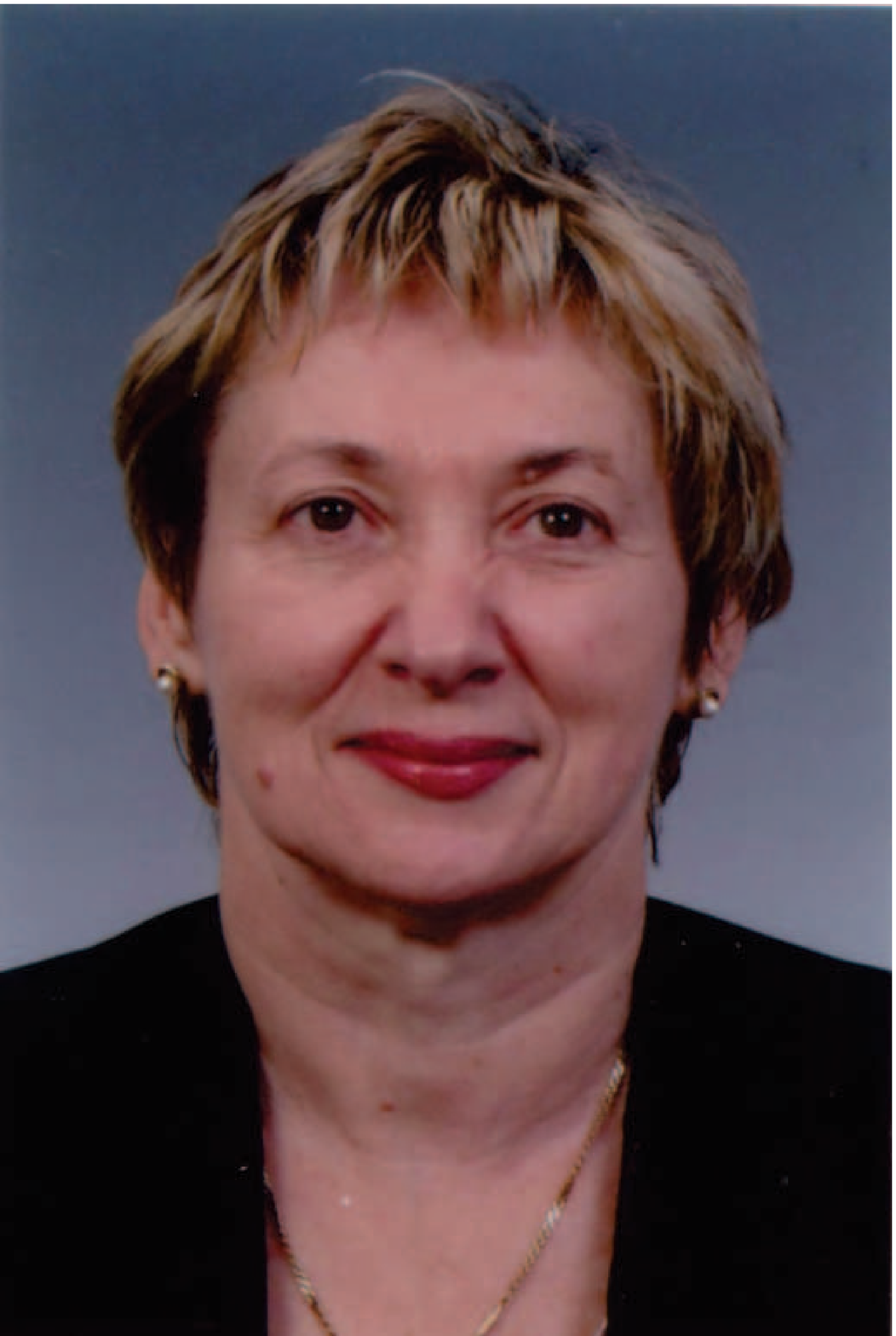}}]{Branka Vucetic}
(M'83-SM'00-F'03) received the
B.S.E.E., M.S.E.E., and Ph.D. degrees in 1972,
1978, and 1982, respectively, in electrical engineering,
from The University of Belgrade, Belgrade,
Yugoslavia. During her career, she has held various
research and academic positions in Yugoslavia,
Australia, and the UK. Since 1986, she has been
with the Sydney University School of Electrical and
Information Engineering in Sydney, Australia. She
is currently the Director of the Centre of Excellence
in Telecommunications at Sydney University. Her
research interests include wireless communications, digital communication
theory, coding, and multi-user detection.

In the past decade, she has been working on a number of industry sponsored
projects in wireless communications and mobile Internet. She has taught a
wide range of undergraduate, postgraduate, and continuing education courses
worldwide. Prof. Vucetic has co-authored four books and more than two
hundred papers in telecommunications journals and conference proceedings.
\end{IEEEbiography}

\end{document}